# Pattern Analysis of World Conflicts over the past 600 years

Gianluca Martelloni[1,3], Francesca Di Patti[1], Ugo Bardi[2*],

[1]Interuniversity Consortium For Science and Technology of Materials (INSTM)

[2]Dipartimento di Chimica, Università di Firenze

Via della Lastruccia 3, 50019 Firenze, Italy

[3]INAF (National Institute for AstroPhysics)

*corresponding author: ugo.bardi@unifi.it

**Abstract**

We analyze the database prepared by Brecke (Brecke 2011) for violent conflict, covering some 600 years of human history. After normalizing the data for the global human population, we find that the number of casualties tends to follow a power law over the whole data series for the period considered, with no evidence of periodicity. We also observe that the number of conflicts, again normalized for the human population, show a decreasing trend as a function of time. Our result agree with previous analyses on this subject and tend to support the idea that war is a statistical phenomenon related to the network structure of the human society.

**Introduction**

The causes of human conflicts remain largely an unresolved subject, especially for the large conflicts that we call "wars." Historians often tend to see wars arising from specific decisions of human actors, in turn the result of specific economic or political strains pitting nations or social groups against each other. But another possible interpretation is that wars are related to the structure of the human society as a whole. Perhaps the first to argue in favor of war as a collective phenomenon was Leon Tolstoy in his *War and Peace* novel (1869) (Emerson 2009). Tolstoy intuition couldn't be verified in quantitative terms until databases about historical wars became available during the 20[th] century. Among the earliest sets of data, we can cite that of Sorokin (Sorokin 1937) and Quincey Wright (Wright 1942). The latter study, published in 1942, listed a total of 284 wars and some 3000 battles from 1500 A.D. to about 1940.

The milestone in this field was the series of studies on "deadly quarrels" published by Lewis Fry Richardson (Richardson L. F. 1960), later discussed in (Wilkinson 1980), (Hess 1995), (Hayes 2002), and (Gallo 2015). Richardson's data cover 315 conflicts, each one involving more than 300 deaths from 1820 to 1949. Richardson himself performed an extensive pattern analysis of the set finding support for the idea that war is a statistical phenomenon embedded throughout human history for the period considered. In particular, Richardson found that, as expected, large wars are less frequent than small ones and that the probability of a war of a certain dimension can be described using a Poisson distribution. Later on, other authors performed similar studies: for instance, Sarkees et al. (Sarkees, Wayman, and Singer 2003) examined a similar time interval, but they didn't perform a statistical pattern analysis. Alvarez-Ramirez et al. (Alvarez-Ramirez et al. 2007) studied the patterns of conflicts during the Iraq war from 2003 onward, finding clear evidence of power laws in the distribution of the military casualties. Gonzales studied a data set of modern wars (González-Val 2016) while Friedman used a reverse approach, that of using power laws to reduce the uncertainty in the historical data (Friedman 2015). Other analyses evidenced the presence of power laws in a variety of datasets involving human conflict, for instance (Bohorquez et al. 2009) and (Johnson et al. 2013). For a review of these studies, see Ward et al., 2013 (Ward et al. 2013) and, more recently. (Clauset 2018), and (Clauset and Gleditsch 2018), confirming previous analyses and discussing the likelihood of a new major conflicts.

Our contribution in this paper is the statistical analysis of the conflict catalog provided by Peter Brecke (Brecke 2011). It is among the largest datasets existing today, reporting data for war fatalities from 1400 to 2000 AD, even though Cirillo and Taleb examined an even longer database, starting from the 1$^{st}$ century AS, although including only relatively large conflicts (Cirillo and Taleb 2015). We examine the trends of the Brecke database, as well as the autocorrelation and the Fourier analysis applied to the time series of fatalities. We consider also the same data normalized to the world population. In the latter case the fatality distribution clearly follows a power law. War seems to follow the same statistical laws as other catastrophic phenomena, such as hurricanes, earthquakes, tsunamis, floods and landslides, whose distribution follows approximate power law (P Bak 1996).

**Materials and method**

Starting from Brecke's database (Brecke 2011), we build the time series that will be statistically analyzed. As a first approach, we examined the number of wars over time. The plot of Brecke's data is shown in figure 1.

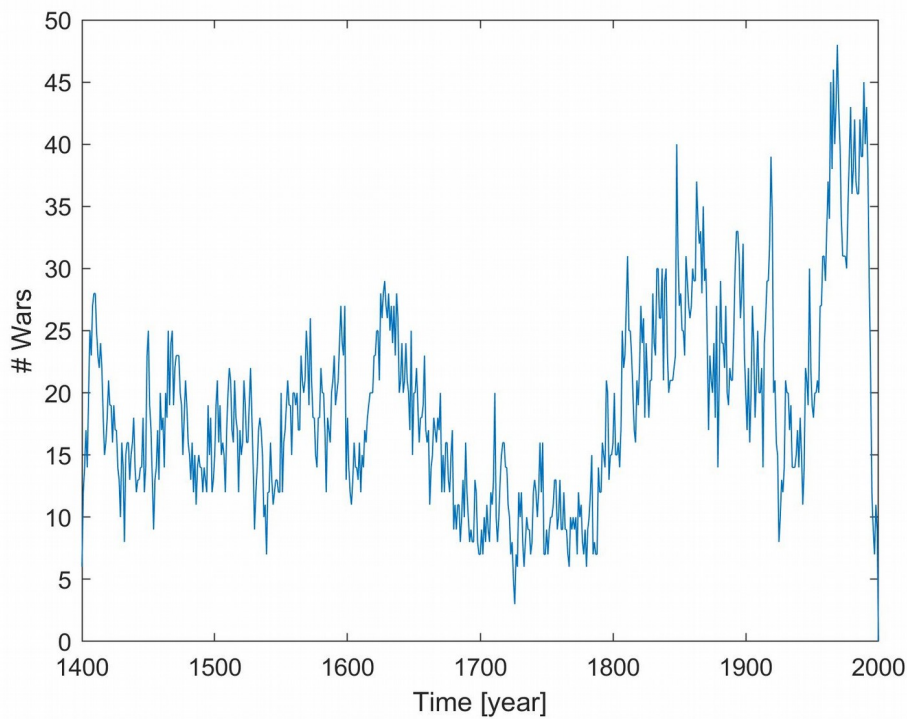

*Figure 1: number of war over time according to Brecke database.*

At a first qualitative examination, the data show an increasing trend, especially noticeable as we get closer to our time, a relatively peaceful period during the 18th century is noticeable. As a further step, we normalized the data for the world population, using the data provided by the History Database of the Global Environment – http://themasites.pbl.nl/tridion/en/themasites/hyde/). In this case (fig. 2) the trend is reversed with the number of wars per unit population clearly decreasing, although the relatively peaceful 18th century remains noticeable.

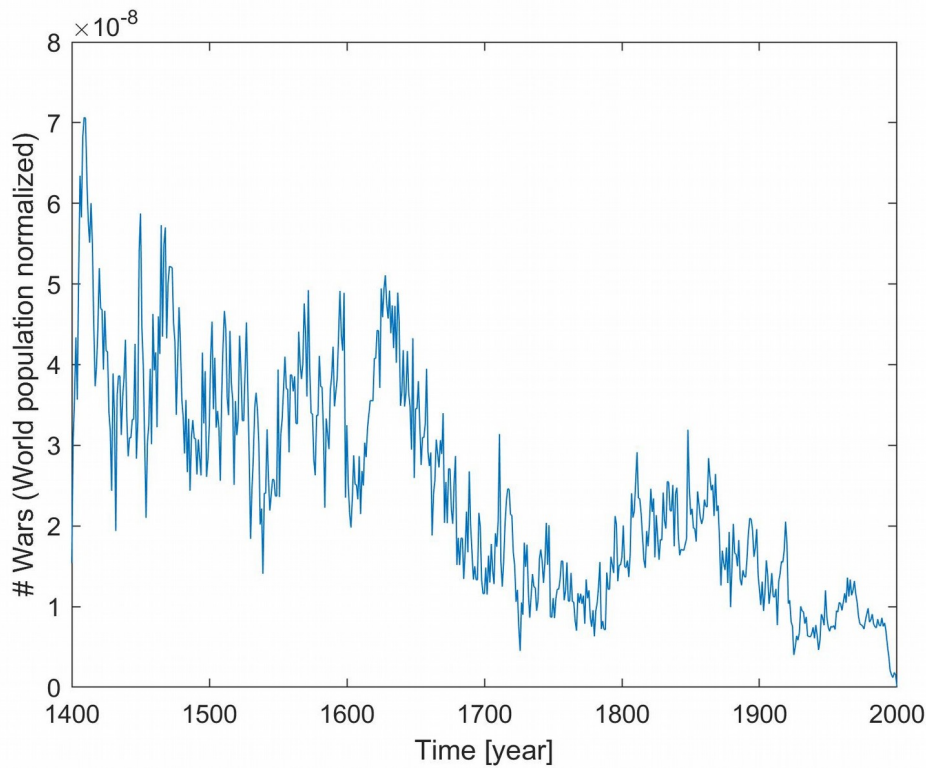

*Figure 2: number of war normalized for the world population.*

In itself, the number of wars is a poor indication of the trends of violent conflicts, hence we proceeded with the analysis of the number of fatalities. Brecke's database provides only the fatalities per war, so we examined the data on the basis of the assumption that, if the duration of a war is greater than one year, the fatalities of that war are summed and collocated in the end war year. The results are reported in figure 3 in terms of total fatalities per year. In figure 4, we show the total fatalities for each war in chronological order. In figures 5-6 we show the same data normalized with respect to to the world population. In these figures, note how there exist years without wars and fatalities (138 out of a total of 601 years).

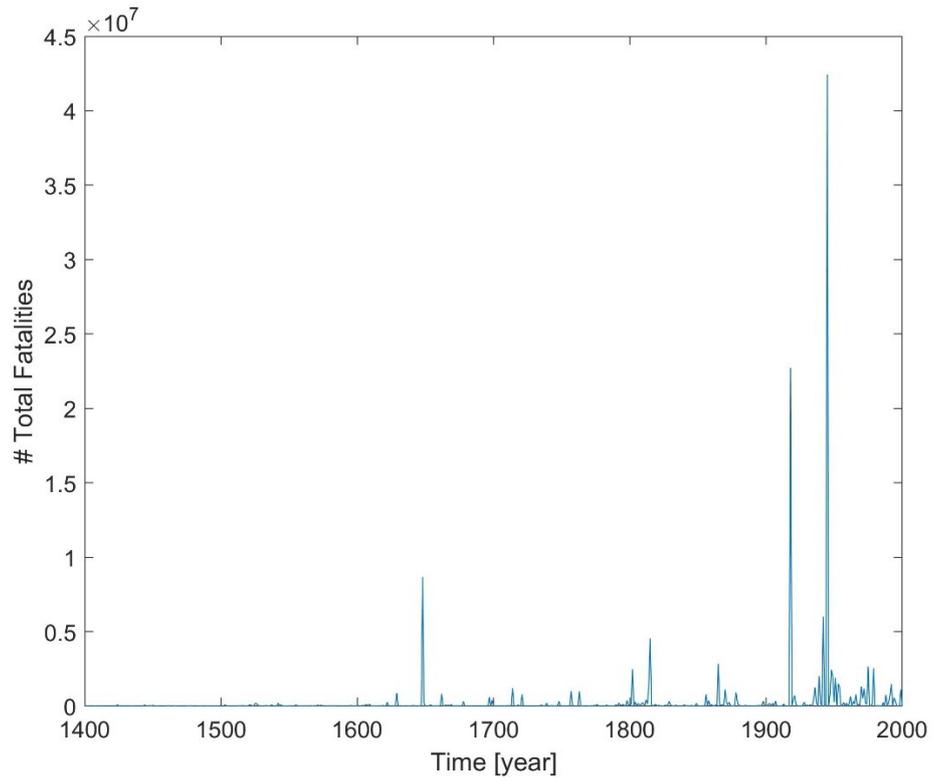

*Figure 3: Total war fatalities per year from Brecke's Database.*

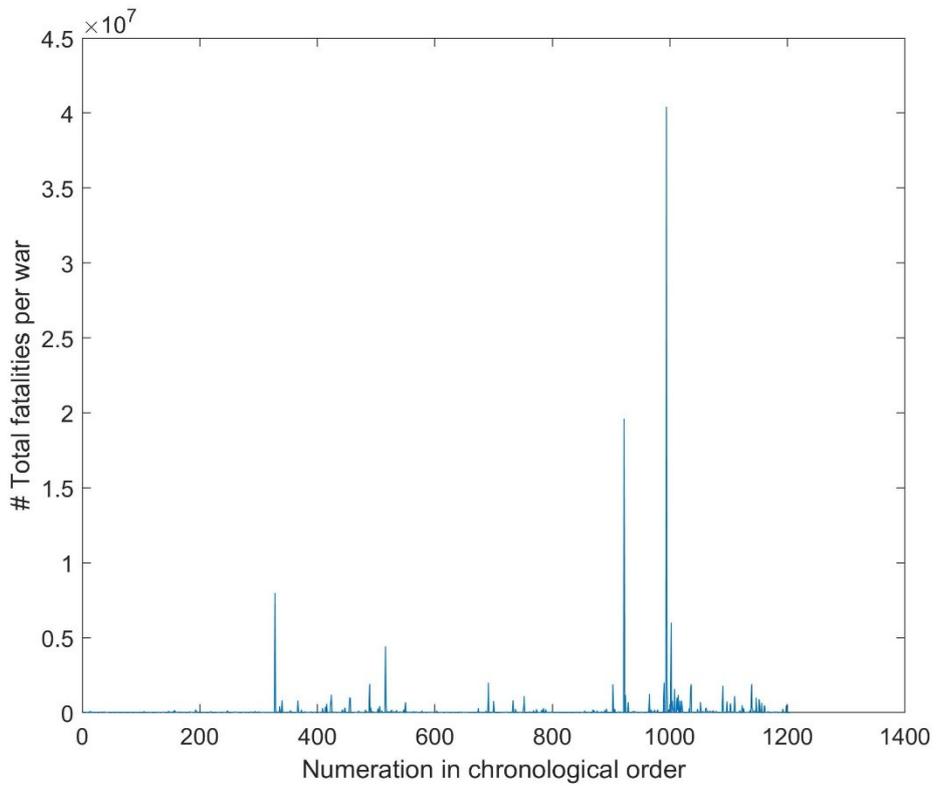

*Figure 4: Total war fatalities per war the numeration is referred to each single war in chronological order.*

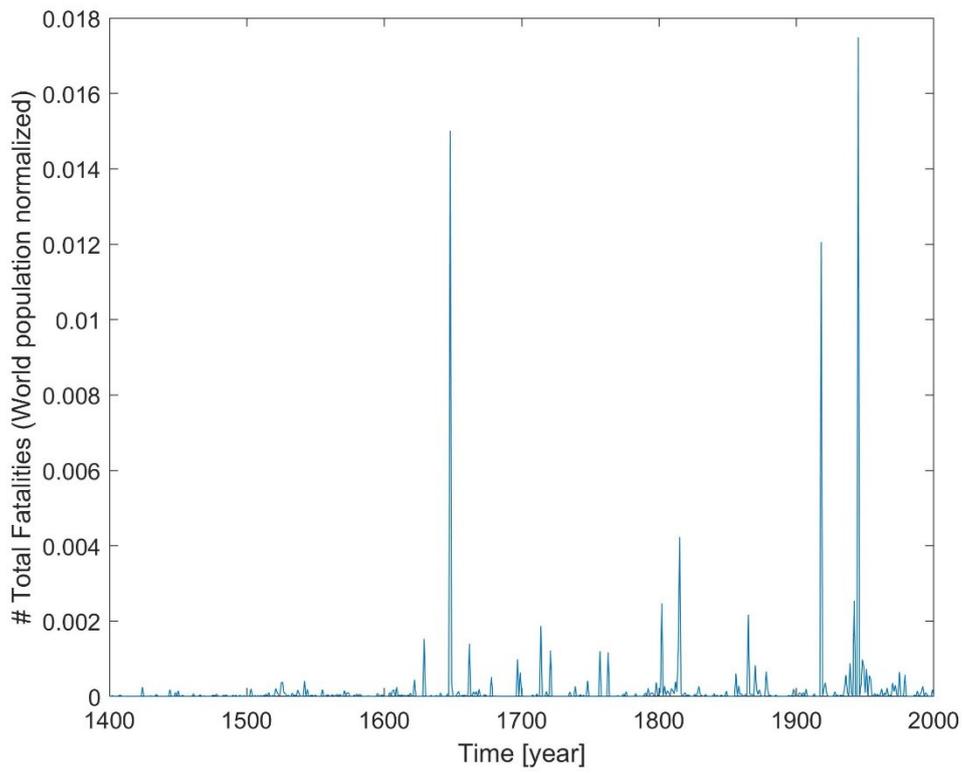

Figure 5: Total war fatalities over time per year normalized to the world population.

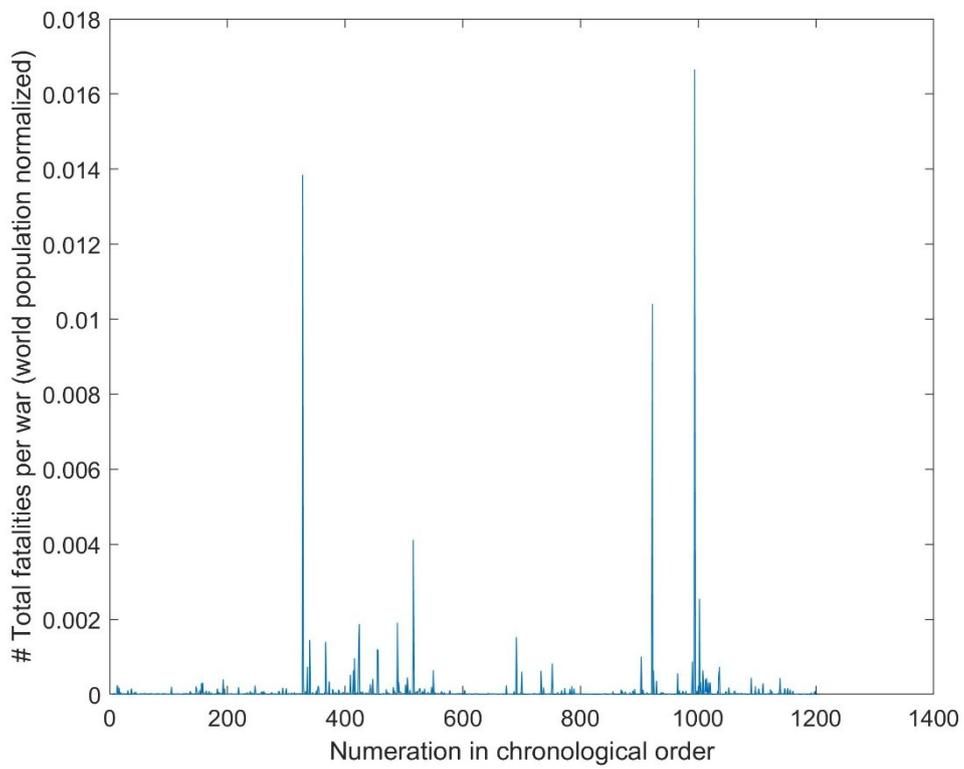

Figure 6: Total war fatalities per war normalized with world population (the numeration is referred to single war in chronological order).

We have therefore four data series: fatalities per year and per conflict, normalized or not for the world's population increase. We analyzed the data in terms of a power law (equation 1) or of a lognormal distribution (equation 2) simply obtained considering the logarithm of data whose corresponding distribution is Gaussian. For the analytical proof of the latter statement, see Martelloni and Bagnoli (2014) (Martelloni and Bagnoli 2014).

$$y = a \cdot x^b \qquad (1)$$

$$y = a_1 * \exp\left(-\left((x-b_1)/c_1\right)^2\right) \qquad (2)$$

The parameters $a$, $b$, $a_1$, $b_1$, and $c_1$ of equations 1-2 can be estimated by means of a fitting procedure. The statistics was completed calculating the autocorrelation function and using the classical Fourier analysis by means of the Power Spectrum on the basis of the Fast Fourier Transform (FFT) algorithm. The autocorrelation for lag $i$ (where $i = 0, 1, …, I$) between the data $y_t$ and $y_{t+i}$ is expressed by equation 3

$$r_i = \frac{c_i}{c_0}, \qquad (3)$$

while $c_0$ is the sample variance of considered time series and $c_i$ is represented by equation 4,

$$c_i = \frac{1}{T} \sum_{t=1}^{T-i} (y_t - \dot{y})(y_{t+i} - \dot{y}) \qquad (4)$$

where $T$ is the length of series. If $q$ is the lag beyond which the theoretical autocorrelation function is effectively zero, the estimated standard error of the autocorrelation at lag $i > q$ (i.e., $q=i - 1$) is

$$SE(r_i) = \sqrt{\frac{1}{T}\left(1 + 2\sum_{k=1}^{q} r_k^2\right)} . \qquad (5)$$

If the time series is completely random the series term of equation 5 tend to zero and therefore the standard error $SE$ approaches to square root of $1/T$. In the result section the autocorrelations are calculated up to the length of the time series, i.e., to $T$-1.

**Results**

If we consider the uncorrected data of the yearly war fatalities (x) from 1400 to 2000 AD and the corresponding probability P(X≥x), plotted every 1000 fatalities, we obtain the empirical distribution of figure 7. The best fit of the distribution is obtained by means of a lognormal function. This result does not exclude a power law distribution, as argued by (González-Val 2016), (Friedman 2015). In particular the central part of the distribution and the tail at the right (the largest conflicts) can be fitted by a power law. In table 1 we report the results of fitting according to figure 7.

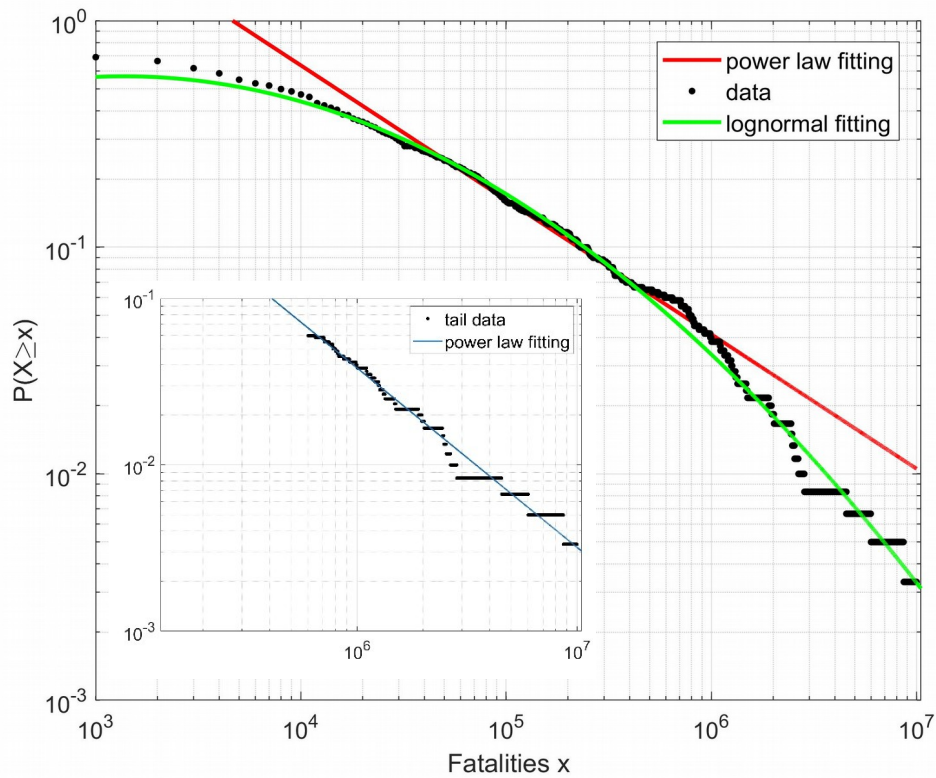

Figure 7: the probability P(X≥x) of war fatalities (per year); the black point marker represents the empirical distribution, the green line is the best fit obtained by means of lognormal function and the red line is a best fit of power law cutting the tails of distribution. In the subfigure we show the power law fit of a tail distribution.

Table 1: results of fatalities distribution P(X≥x) fitting.

| Distribution | Lognormal (figure 7) | Power law red line (figure 7) | Power law right tail (subfigure of figure 7) |
|---|---|---|---|
| SSE | 0.002619 | 0.0026 | 0.004094 |
| $R^2$ | 0.9998 | 0.9995 | 0.9963 |
| Adjusted- $R^2$ | 0.9998 | 0.9995 | 0.9963 |
| RMSE | 0.0005118 | 0.001321 | 0.0006599 |
| $a_1$ | 0.5686 | | |
| $b_1$ | 7.225 | | |
| $c_1$ | 3.919 | | |
| $a$ | | 150.5 | $1.149 \cdot 10^5$ |
| $b$ | | -0.5937 | -1.08 |

Completing the investigation, the autocorrelation and the Fourier analysis show that the time series of fatalities is completely random (see figures 8-9). That is, no periodicity is detectable in this time series.

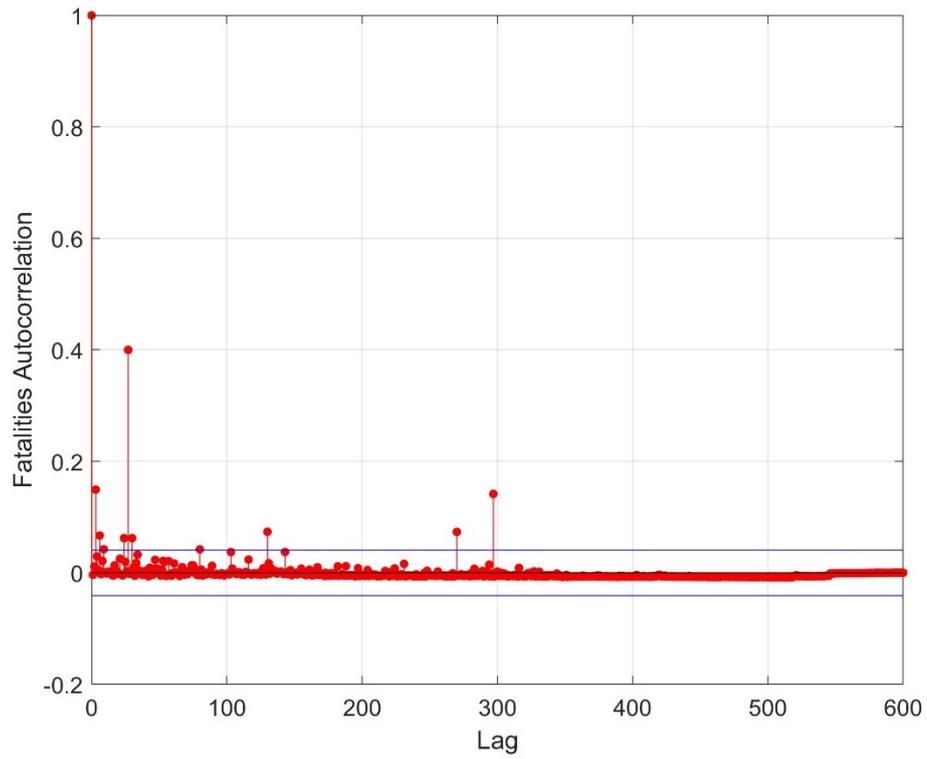

*Figure 8: the fatalities autocorrelation; the two blue lines represent the standard error of the autocorrelation (the value is 0.0408 equal to the root square of 1/T that indicates the series is completely random).*

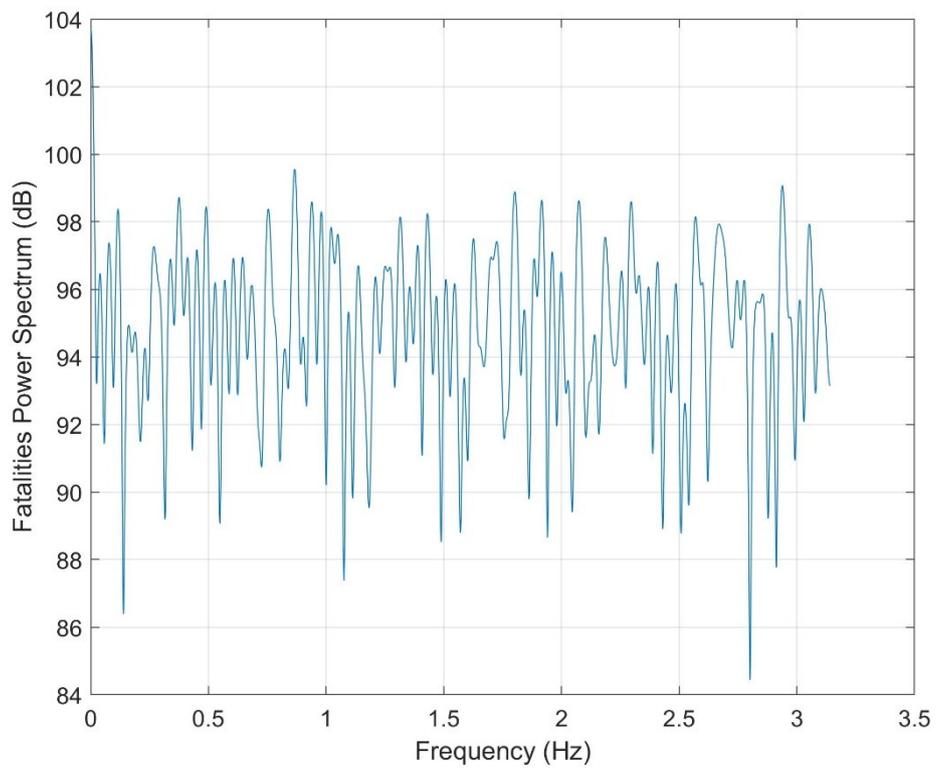

*Figure 9: analysis of time series of war fatalities by means of Power Spectrum (by means of FFT).*

We can draw similar conclusions if we consider the time series of fatalities per war (see figure 10). In table 2 we report the results of fitting according to figure 10. We do not report the corresponding autocorrelation function and power spectrum as these are similar to the results shown before. In this case the standard error of the autocorrelation is 0.0288, equal to the root square of 1/T that indicates the series is completely random. Again, no periodicity is detectable in these data.

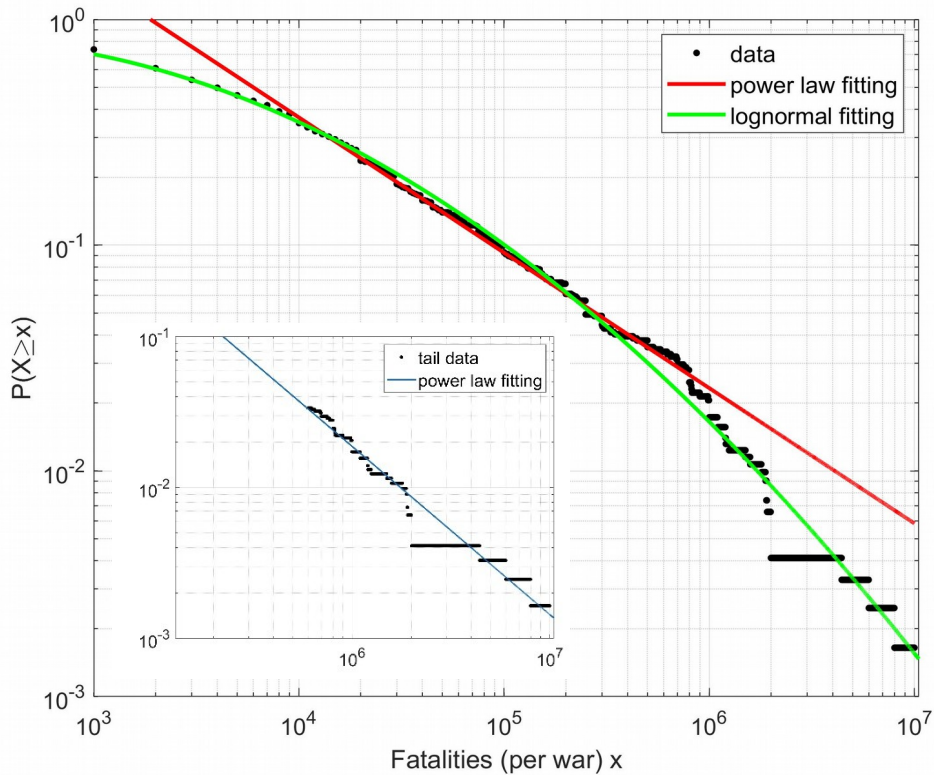

Figure 10: the probability $P(X \geq x)$ of war fatalities (per war); the black point marker represents the empirical distribution, the green line is the best fit obtained by means of lognormal function and the red line is a best fit of power law cutting the tails of distribution. In the subfigure we show the power law fit of a tail distribution.

Table 2: results of fatalities per war distribution $P(X \geq x)$ fitting.

| Distribution | Lognormal (figure 10) | Power law red line (figure 10) | Power law right tail (subfigure of figure 10) |
|---|---|---|---|
| SSE | 0.0009828 | 0.0042 | 0.0008289 |
| $R^2$ | 0.9998 | 0.999 | 0.9971 |
| Adjusted- $R^2$ | 0.9998 | 0.999 | 0.9971 |
| RMSE | 0.0003135 | 0.001674 | 0.0002969 |
| $a_1$ | 0.8185 | | |
| $b_1$ | 5.195 | | |
| $c_1$ | 4.363 | | |
| $a$ | | 92.23 | $9.306 \cdot 10^4$ |
| $b$ | | -0.5997 | -1.116 |

A further step in the analysis is to examine the data after normalizing for the world population. In this case the distributions clearly follow a power law, not a lognormal one. The normalization impacts mainly on the left tail, that is magnifying the importance of ancient, relatively small wars. The results are shown in figures 11-12.

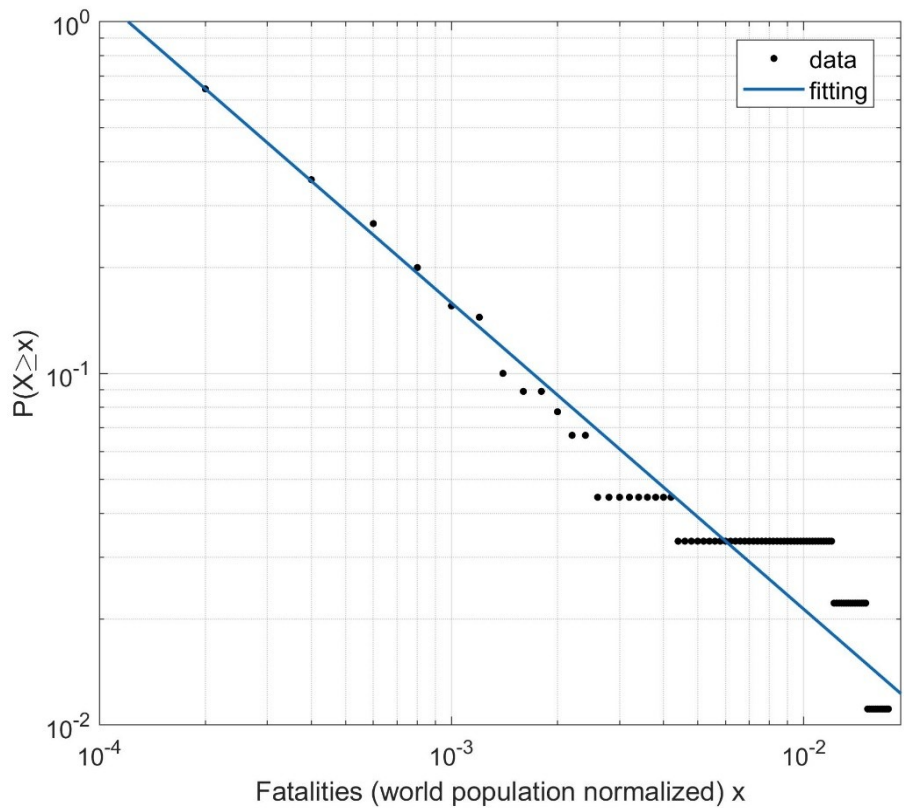

*Figure 11: Distribution of war fatalities world population normalized fitted by power law (equation 1) - a=0.0003873, b=-0.8708, SSE=0.007736, $R^2$=0.9869, Adjusted- $R^2$=0.9868, RMSE=0.009376.*

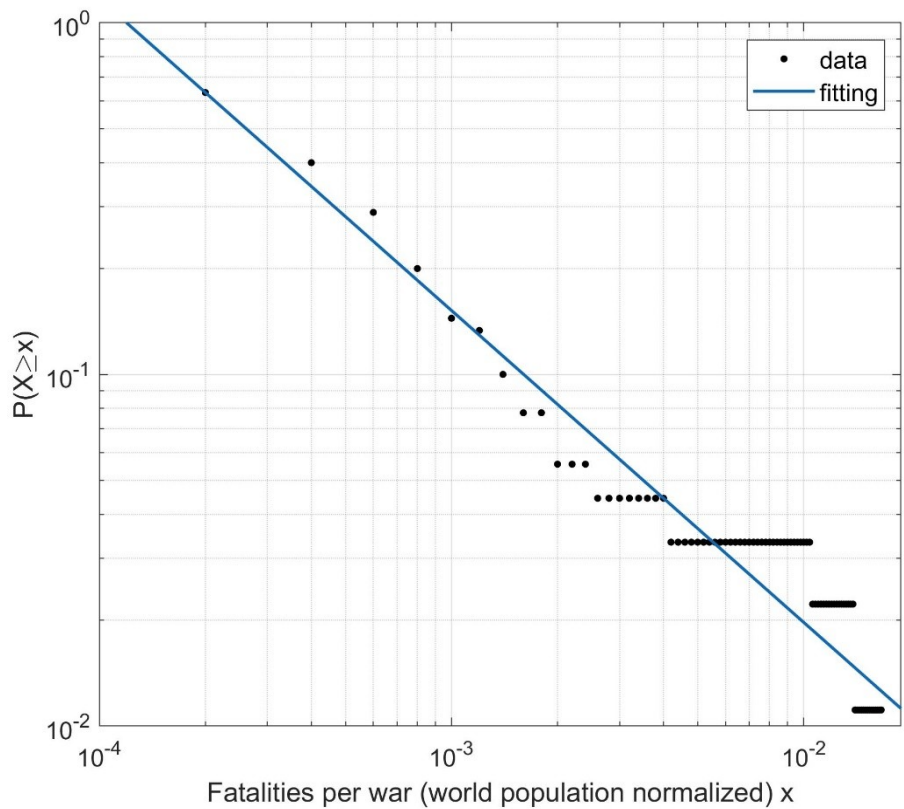

*Figure 12: Distribution of war fatalities per war world population normalized fitted by power law (equation 1) - a=0.000332, b=-0.8869, SSE=0.0005404, $R^2$=0.9991, Adjusted- $R^2$=0.9991, RMSE=0.002478.*

Also, in these cases, the autocorrelation and the Fourier analysis show that the time series of fatalities is completely random. The data are not reported here since the results are similar to those obtained for the non-normalized time series. The standard error of the autocorrelation is 0.0408 for war fatalities world population normalized and 0.0288 for war fatalities per war world population normalized, equal to the root square of 1/T that indicates the series is completely random.

**Conclusions**

Our contribution in this field consists in validating the Brecke conclict database (Brecke 2011), among the longest and most complete ones available. This analysis confirms previous work, see e.g. (Clauset 2018), (Clauset and Gleditsch 2018) for a general discussion. The data indicate that power laws are common in the distribution of violent conflicts in human history: in this case, the trend is clear when the number of casualties are normalized for the increasing world population. Note also that the normalized number of conflicts per year tend to decrease with time – this result indicates that in modern times wars have tended to become less frequent but more destructive. In practice, these result confirm that there is little evidence supporting the idea popularized by Pinker (Pinker 2011) that humankind is progressing toward a more peaceful world. A new major conflict might be possible in a non-remote future, as discussed among others by Clauset (Clauset 2018). These result seem to indicate that human conflicts are a critical phenomenon: we could say that humans worldwide tend to form societies existing in a self-organized critical condition, as defined by Bak et al. (Per Bak, Tang, and Wiesenfeld 1988). In these conditions, war is simply one of the methods that the system has to dissipate entropy at the fastest possible speed (Kleidon, Malhi, and Cox 2010),(Trinn 2018). In other words, war appears to be an unavoidable consequence of the behavior of human beings, and perhaps of other primate species (de Waal 2000). On the other hand, we need also to remark that the available data series involves only a small fraction of human history, corresponding to a period of tumultuous and rapid expansion of both the economy and the population. A future of declining natural resources might show different trends.

Acknowledgement: the authors are grateful to Aaron Clauset for his comments and suggestions.